\documentclass[prl, twocolumn, showpacs, amsmath,amssymb, floatfix, eqsecnum]{revtex4-1}
\pdfoutput=1
\usepackage{amsmath}
\usepackage{amssymb}
\usepackage{amsthm}
\usepackage{amsfonts}
\usepackage{comment}

\usepackage{graphicx,color,framed}
\usepackage{hyperref}
\usepackage{times}
\usepackage{enumerate}
\usepackage{lipsum}
\usepackage{slashed}
\usepackage{url}
\usepackage{bbm}
\usepackage{chngcntr}
\counterwithout{equation}{section}

\hypersetup{
    colorlinks=true, 
    linktoc=all,     
    linkcolor=blue,  
}

\def \beq {\begin{equation}}
\def \eeq {\end{equation}}
\def \beqa {\begin{eqnarray}}
\def \eeqa {\end{eqnarray}}
\def \bseq {\begin{subequations}}
\def \eseq {\end{subequations}}

\newcommand{\<}{\langle}
\renewcommand{\>}{\rangle}

\usepackage{soul}
\definecolor{DarkRed}{RGB}{100,0,0}

\definecolor{DarkGreen}{RGB}{0,100,0}

\newtheorem{theorem}{Theorem}

\newtheorem{lemma}{Lemma}

\begin{document}

\title{Vanishing Hall conductance for commuting Hamiltonians}

\author{Carolyn Zhang}\email{cczhang@uchicago.edu}
\author{Michael Levin}\email{malevin@uchicago.edu}
\affiliation{Department of Physics, Kadanoff Center for Theoretical Physics, University of Chicago, Chicago, Illinois 60637,  USA}
\author{Sven Bachmann}\email{sbach@math.ubc.ca}
\affiliation{Department of Mathematics, The University of British Columbia, Vancouver BC V6T 1Z2, Canada}


\begin{abstract}
We consider the process of flux insertion for ground states of almost local commuting projector Hamiltonians in two spatial dimensions. In the case of finite dimensional local Hilbert spaces, we prove that this process cannot pump any charge and we conclude that the Hall conductance must vanish.
\end{abstract}

\maketitle

\textbf{\emph{Introduction}}.--- A local commuting projector Hamiltonian (LCPH) is a special kind of quantum lattice model of the form $H = \sum_r H_r$, where each $H_r$ is a projection operator supported on a finite collection of nearby lattice sites, and where the different $H_r$'s commute with one another. Lattice models of this kind, such as the toric code model\cite{Kitaev2003fault}, have proven to be  
powerful tools for studying interacting topological phases of matter. 
Given the many applications of these models\cite{levin2005,chen2013,heinrich2016,cheng2017,haah2011}, it is important to understand their limitations: that is, what phases \emph{cannot} be realized by LCPHs? In two dimensions, it is known\footnote{To see why the thermal Hall conductance must vanish for commuting Hamiltonians, note that the energy current $f_{jk}=0$ in Eq.~(154) of Ref.~\onlinecite{kitaev2006anyons}, and hence one can choose $h_{jkl} = 0$ in Eq.~(159), which leads to a vanishing thermal Hall conductance in Eq.~(160).} that one class of such phases are those with a nonzero thermal Hall conductance\cite{kitaev2006anyons}. In this work, we show that another class of such phases are those with a nonzero \emph{electric} Hall conductance $\nu$. 
This claim was first proved in Ref.~\onlinecite{kapustin2020} using algebraic geometry. Here, we give a simple and physically motivated proof based on the idea of flux insertion. Our techniques are closely related to those of Refs.~\onlinecite{bachmann2020,bachmann2021,kapustin2020hall}. Our argument has the additional advantage that it extends to \emph{almost} local CPHs (ALCPHs), a generalization of LCPHs that includes Hamiltonians with interactions that decay faster than any power.

\textbf{\emph{Physical argument.}}---We first present an intuitive, but non-rigorous, argument for our no-go result. This argument is similar to our main argument, but not as general, since it applies only to \emph{strictly} local commuting projector Hamiltonians. It also assumes the ``local topological quantum order'' (LTQO) property (\ref{ltqo}), which is stronger than the property (\ref{wTQO}) used in the main argument. 

\begin{figure}[tb]
   \centering
   \includegraphics[width=.8\columnwidth]{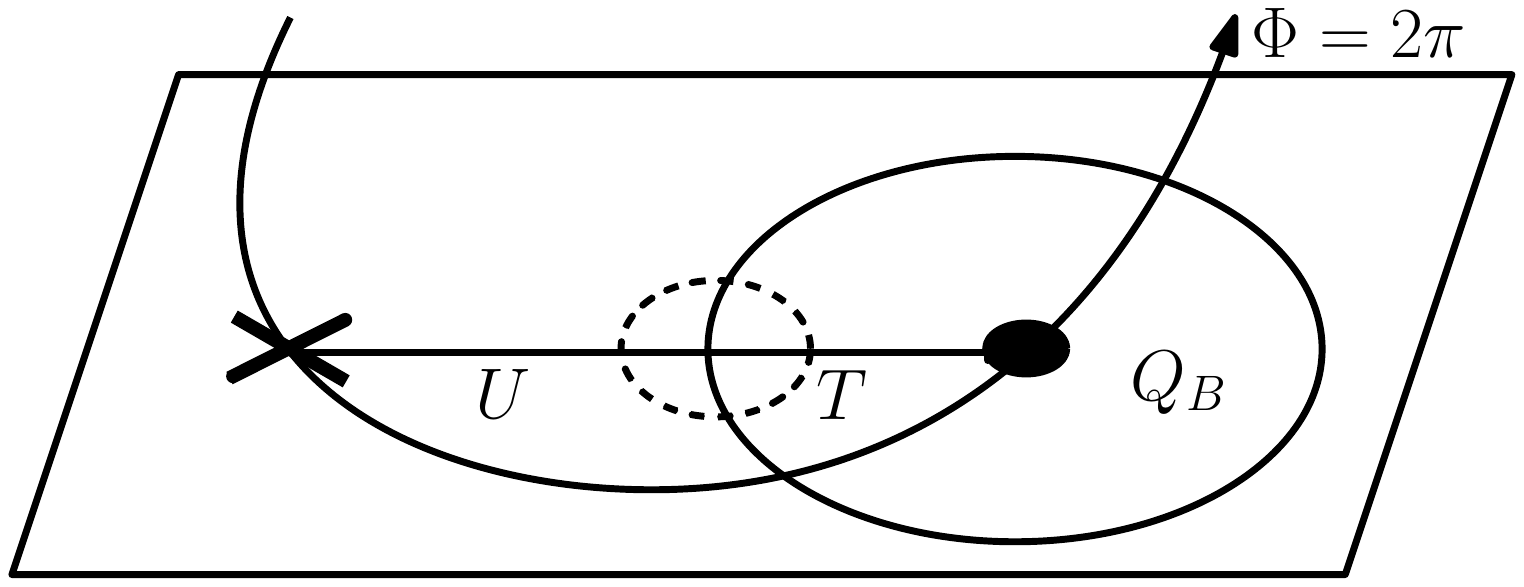} 
   \caption{Setup for physical argument. A string operator $U$ inserts $\pm 2\pi$ flux at its endpoints. This operation pumps charge from one puncture to the other, increasing the total charge $Q_B$ within region $B$ (solid circle). The operator $T = U^\dagger Q_B U - Q_B$ measuring the change in $Q_B$ is supported in the small dotted circle. For an LCPH, $U$ commutes exactly with all Hamiltonian terms that are supported away from the two punctures.}
   \label{fig:fluxstring}
\end{figure}

Imagine starting in a ground state $|\Omega\>$ of a two dimensional LCPH and then adiabatically inserting $\pm 2\pi$ flux at two punctures. This process can be implemented by a string operator $U$ localized along a line between the two punctures, as illustrated in Fig.~\ref{fig:fluxstring}. By the Laughlin argument~\cite{laughlin1981quantized,avron1994charge}, the amount of charge pumped by this process from one puncture to the other is equal to $2\pi\nu$. Let $B$ be a region surrounding one of the two punctures, and let $Q_B$ be the operator that measures the total charge in region $B$. The charge pumped by the flux insertion is then $\<\Omega| U^\dagger Q_BU-Q_B | \Omega \>$, so the Hall conductance is
\begin{equation}\label{hallintro}
    \nu = \frac{1}{2\pi}\< \Omega| U^\dagger Q_BU-Q_B | \Omega \>,
\end{equation}
Since the system is charge conserving and the current flows only along the string, the operator $T \equiv U^\dagger Q_BU-Q_B$ is localized near the point where the string intersects the boundary of $B$, as indicated in Fig.~\ref{fig:fluxstring}.

Now consider the charge pumped by inserting many units of flux, written as a telescoping sum:
\begin{equation}\label{hallintro2}
    \<\Omega| U^{\dagger n}Q_BU^n-Q_B |\Omega\> =\sum_{k=0}^{n-1}\< \Omega| U^{\dagger k}TU^k | \Omega \>.
\end{equation}
Crucially, for an LCPH, the operator $U$ commutes exactly with all $H_r$ terms away from the two punctures (we justify this claim below). This means, in particular, that $U^k |\Omega\>$ does not contain any excitations away from the two punctures, i.e.~it is a ``local ground state'' in this region. Then, assuming that the Hamiltonian obeys the local TQO condition (\ref{ltqo}), we deduce that $U^k |\Omega\>$ must have the same expectation values as $|\Omega\>$ for any local observable supported away from the punctures. In particular, specializing to the observable $T$, we deduce that $\<\Omega | U^{k \dagger} T U^k | \Omega\> = \<\Omega | T | \Omega\>$. We conclude that all of the terms in the sum in (\ref{hallintro2}) give the same quantity $2\pi \nu$, so the right hand side evaluates to $2\pi n \nu$. 

At the same time, the absolute value of the left hand side of (\ref{hallintro2}) is bounded by $ |q_{\mathrm{max}} - q_{\mathrm{min}}|$ where $q_{\mathrm{max}}$ and $q_{\mathrm{min}}$ are the largest and smallest eigenvalues of $Q_B$. 
Hence, we have the bound $2\pi n |\nu| \leq |q_{\mathrm{max}} - q_{\mathrm{min}}|$. Since $n$ can be made arbitrarily large, we conclude that $\nu = 0$.

To complete the argument, we now explain why $U$ commutes with all the $H_r$ terms away from the two punctures. The key point is that all the $H_r$ terms that are supported away from the two punctures remain commuting projectors throughout the flux insertion process\cite{kapustin2020}. Therefore, if the system starts in an eigenspace of some $H_r$ away from the punctures, it will stay in this eigenspace throughout the flux insertion, since all the terms in the Hamiltonian commute with $H_r(t)$ at all times, and the process is adiabatic. In particular, the system is in the same eigenspace at the end of the process as at the beginning, implying that $U$ commutes with $H_r$.

This behavior should be contrasted with that of non-commuting Hamiltonians: in that case, $U$ does not commute with $H_r$, so there is no reason that $U^k|\Omega\>$ has to be in a ground state away from the two punctures for arbitrarily large $k$. Instead, every time we apply $U$, we create additional (possibly charged) excitations, that spread outward from the two punctures as $k$ increases. When $k$ is large enough, the excited region in $U^k |\Omega\>$ reaches the support of $T$ and the total pumped charge stops growing linearly. Hence $\nu$ can be nonzero without contradiction.

An important loophole in the above no-go argument is that it assumes that $Q_B$ has a bounded spectrum. This assumption can break down if the Hilbert space on each lattice site is infinite dimensional. This explains how the LCPH in Ref.~\onlinecite{demarco2021} can realize a state with $\nu \neq 0$: the example given there uses a system built out of infinite dimensional rotor degrees of freedom. In such a system, a finite region $B$ can absorb an infinite amount of charge, so $\nu$ can be nonzero.



We now turn to a rigorous version of this argument based on \emph{infinitesimal} flux insertion. This argument applies to a more general class of almost local commuting projector Hamiltonians (ALCPHs).

\textbf{\emph{Setup}}---We consider a sequence of two dimensional lattice spin systems of increasing linear size $L$, defined on a torus geometry. We denote the lattice by $\Lambda=\{-L/2+1,\cdots,L/2-1,L/2\}^2$ where we take $L$ to be even for convenience. Each site $r\in\Lambda$ corresponds to a finite dimensional local Hilbert space, where the dimension is fixed and does not depend on $L$. In the following, all constants are uniform in the system size; the notation $\mathcal{O}(L^{-\infty})$ means $\leq C_k L^{-k}$ for all $k$, for some constant $C_k$.

We consider Hamiltonians that are sums of commuting projectors of the form
\begin{equation}\label{ALCPH1}
    H=\sum_{r\in\Lambda} H_r, \quad [H_r,H_{r'}]=0,\quad H_r^2=H_r=H_r^\dagger.
\end{equation}
Each projector $H_r$ is ``almost local'' in the sense that $H_r$ commutes with operators $O_{r'}$, supported on a single site $r'$, up to error superpolynomially small in the distance $|r-r'|$:
\begin{equation}\label{ALCPH3}
    \|[H_r,O_{r'}]\|\leq\|O_{r'}\|\cdot\mathcal{O}(|r-r'|^{-\infty}).
\end{equation}
We also assume each $H_r$ is charge conserving:
\begin{equation}
    [H_r, Q_\Lambda]=0\qquad\forall r\in\Lambda,
\end{equation}
where $Q_{\Lambda}=\sum_{r\in\Lambda}q_r$ is a sum of Hermitian charge operators $q_r$, each supported on site $r$, with an integer spectrum and a uniformly bounded norm. In addition, we assume that the number of ground states of $H$ remains bounded as $L \rightarrow \infty$ and that these ground states are simultaneous eigenstates of the projectors $\{H_r: r \in \Lambda\}$ with eigenvalue $0$ (i.e.~$H$ is frustration-free). 

To state our final assumption, we first need to introduce some notation. For any region $R$, we define the corresponding ``local ground state subspace'' $V_R$ to be the set of all states that are annihilated by the projectors $\{H_r:r\in R\}$. We denote the projector onto $V_R$ by $P_R$, and we denote the expectation value of an observable $O$, averaged over $V_R$, by $\langle O\rangle_{R}=\frac{1}{\mathrm{Tr}(P_{R})}\mathrm{Tr}(P_{R}O)$. We use the abbreviation $P \equiv P_\Lambda$ to denote the projector onto the global ground state subspace, and likewise we use the notation $\<O\> \equiv \<O\>_\Lambda$ to denote the average over the global ground state subspace.

Our final assumption is a weaker version of the local topological order (LTQO) condition of Refs.~\onlinecite{bravyi2010,michalakis2013}. The usual LTQO condition states that for any region $R$, and any local observable $O_{\tilde R}$ supported in a smaller region $\tilde{R} \subset R$, the expectation value of $O_{\tilde R}$ in (any) local ground state $|\Psi_R\> \in V_R$ is approximately the same as the expectation value in (any) global ground state $|\Omega\>$:
\begin{equation}
    \<\Psi_R| O_{\tilde{R}} |\Psi_R\> = \<\Omega| O_{\tilde{R}}|\Omega\> + \|O_{\tilde{R}}\|\cdot\mathcal{O}(\mathrm{dist}(\tilde R, R^c)^{-\infty}),
\label{ltqo}
\end{equation}
where $R^c$ is the complement of $R$ in $\Lambda$. 
In this paper, we will only need the weaker property that the \emph{average} expectation value of $O_{\tilde R}$ over the local ground state subspace $V_R$ is approximately equal to the average expectation value of $O_{\tilde R}$ over the global ground state subspace:
\begin{equation}\label{wTQO}
    \langle O_{\tilde R}\rangle_{R} = \langle O_{\tilde R}\rangle + \|O_{\tilde R}\|\cdot\mathcal{O}(\mathrm{dist}(\tilde R, R^c)^{-\infty}).
\end{equation}

Note that (\ref{wTQO}), unlike (\ref{ltqo}), does not require local indistinguishability of ground states. Rather, it can be interpreted as a local response condition: it says that local observables $O_{\tilde R}$ have approximately the same (zero temperature) expectation values in the full system as they do in a subsystem $R \supset \tilde{R}$. One difference between (\ref{wTQO}) and the usual LTQO condition (\ref{ltqo}) is that (\ref{wTQO}) can be satisfied by systems with spontaneous symmetry breaking, while such systems generally violate (\ref{ltqo}).



\begin{figure}[tb]
   \centering
   \includegraphics[width=.8\columnwidth]{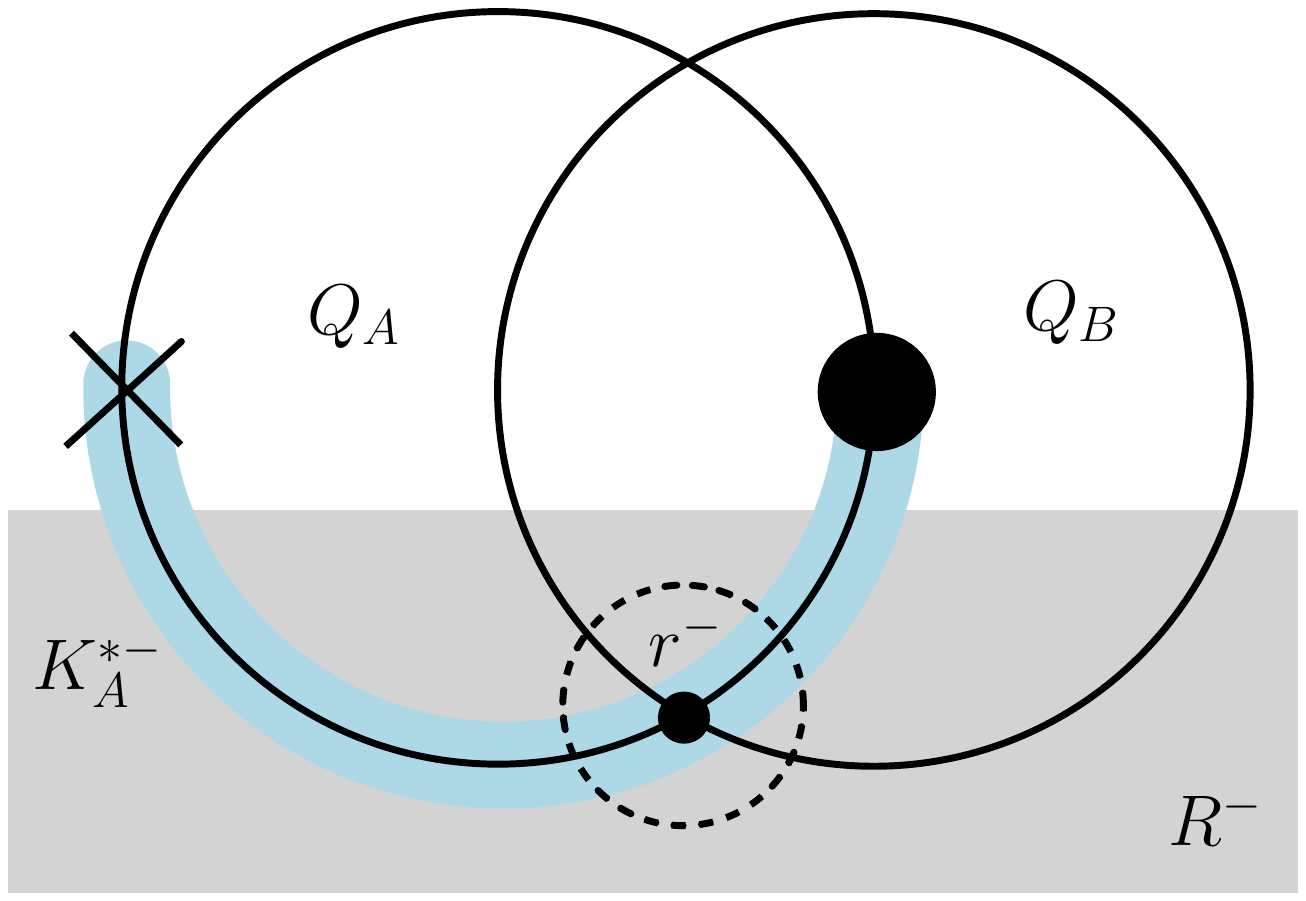} 
   \caption{Geometry of main argument. Two disks $A, B$ with charge $Q_A, Q_B$, intersect in the lower half torus at point $r^-$. The operators $K_A^{*-}$ and $[K_A^{*-},Q_B]$ are supported in the blue strip and the dotted circle respectively. The operator $Q_A - K_A^{*-}$ commutes (up to $\mathcal{O}(L^{-\infty})$) with $P_{R^-}$, the projector into the local ground state subspace of the (shaded) region $R^-$.}
   \label{fig:hallcond}
\end{figure}
\textbf{\emph{Hall conductance}}.---To define the Hall conductance within this setting, we consider a geometry consisting of two overlapping disks $A$ and $B$ of radius $\frac{L}{4}$, centered at $(-\frac{L}{8},0)$ and $(\frac{L}{8},0)$ respectively (see Fig.~\ref{fig:hallcond}). 
 
Our definition involves a string operator $K_A^-$ that runs along the lower half boundary of $A$ and that inserts an infinitesimal flux into the center of $B$.
To construct $K_A^-$, we assume the existence of an operator $K_A$ with two properties.
First, $K_A$ satisfies
\begin{equation}\label{KR1}
    [Q_A-K_A,P]=0.
\end{equation}
Second, $K_A$ is supported ``near'' the boundary of $A$. More precisely, $K_A$ can be approximated, up to $\mathcal{O}(L^{-\infty})$, by a sum of terms of the form
\begin{equation}\label{KR2}
K_A = \sum_{r\in\partial_{\alpha L}A}\overline{K}_{r,A}+\mathcal{O}(L^{-\infty})
\end{equation} 
where $\partial_{\alpha L}A=\{r\in\Lambda:\max\left(\mathrm{dist}(r,A),\mathrm{dist}(r,A_c)\right)\leq \alpha L\}$ is a strip of width $2\alpha L$ along the boundary of $A$, with $0<\alpha\leq\frac{1}{32}$. Here, each $\overline{K}_{r,A}$ is a strictly local charge conserving operator, with a uniformly bounded norm, supported in $D_{\alpha L}(r)$, a disk of radius $\alpha L$ centered at $r$. It has been shown that an operator $K_A$ with these two properties can be constructed for all gapped, charge conserving Hamiltonians\cite{hastings2015,bachmann2020}. 


Given a $K_A$ with these properties, we construct a corresponding string operator $K_A^-$ by restricting the sum in (\ref{KR2}) to sites in the lower half torus, which we denote by $\Lambda^-=\{r\in \Lambda:r_y\leq 0\}$: 
\begin{equation}\label{kmindef}
    K_A^-=\sum_{r\in(\partial_{\alpha L}A)\cap \Lambda^-}\overline{K}_{r,A}.
\end{equation}

To see why $K_A^-$ inserts an infinitesimal flux, note that for $\theta \ll 1$, the operator $e^{i \theta K_A}$ has the same action on ground states as the gauge transformation $e^{i \theta Q_A}$ by (\ref{KR1}); likewise, the restricted operator $e^{i \theta K_A^-}$ acts like a gauge transformation along the lower boundary of $A$ but acts trivially along the upper boundary of $A$, exactly as one expects for an infinitesimal flux insertion operator.

With these preliminaries, we can now define the Hall conductance in a form that is most convenient for our purposes:
\begin{equation}\label{hallcond}
    \nu=-i\lim_{L\to\infty}\langle[K_A^-,Q_B]\rangle.
\end{equation}
This expression can be interpreted as the charge pumped into $B$ by an infinitesimal flux insertion. 

Note that (\ref{hallcond}) can be related to the more familiar Kubo formula for the Hall conductance. Using~(\ref{KR1}) but with $B$ instead of $A$, we see that $\langle[K_A^-,Q_B]\rangle = \langle[K_A^-,K_B]\rangle$ by cyclicity of the trace. $K_B$ can then be replaced by $K_B^-$, up to $\mathcal{O}(L^{-\infty})$, giving $\nu=-i\lim_{L\to\infty}\langle[K_A^-,K_B^-]\rangle$. This is the adiabatic curvature~\cite{hastings2015,bachmann2018}, which is well-known \cite{avron1985quantization} to express the Kubo linear response coefficient in the QHE.


Importantly, \emph{any} $K_A$ satisfying (\ref{KR1}) and (\ref{KR2}) is valid for computing $\nu$. We will leverage this non-uniqueness of $K_A$ in this work, by constructing a $K_A$ with special properties.

\textbf{\emph{Main result}}.---We now use (\ref{hallcond}) to compute the Hall conductance for ALCPHs. Our main result is the following:
\begin{theorem}
\label{mainresult}
Let $H$ be a charge conserving ALCPH. There is a choice of $K_A$, which we call $K_A^{*}$, satisfying (\ref{KR1},\ref{KR2}) such that the corresponding operator $K_A^{*-}$, defined as in (\ref{kmindef}), obeys
\begin{equation}\label{Thm}
    \langle[K_A^{*-},Q_B]\rangle=\mathcal{O}(L^{-\infty}).
\end{equation}
In particular, $\nu=0$. 
\end{theorem}

\emph{Proof.} Let $K_A^*$ be defined by
\begin{equation}\label{Kdef}
K_A^{*}\\
= Q_A - \int\!\mathcal{D}_\Lambda[\theta] e^{i(\theta,H)}Q_Ae^{-i(\theta,H)},
\end{equation}
where $\mathcal{D}_\Lambda[\theta] = \prod_{r\in\Lambda}\frac{d\theta_r}{2\pi}$, $(\theta,H) = \sum_{r\in\Lambda} \theta_r H_r$, and we integrate over $\{0\leq\theta_r\leq 2\pi\}$. With this definition, the operator $Q_A - K_A^{*}$ is simply an \emph{average} of $U Q_A U^\dagger$ 
over all unitary operators $U = e^{i (\theta, H)}$ generated by the commuting projectors, $H_r$. Therefore, by construction, $Q_A - K_A^{*}$ commutes with every unitary operator $e^{i (\theta, H)}$, and hence it also commutes with the generators, $H_r$: 
%
%
\begin{equation}\label{commH}
[Q_A-K_A^{*},H_r]=0, \quad(r\in\Lambda)
\end{equation}
This ensures that $K_A^*$ satisfies (\ref{KR1}). 


In fact, $K_A^*$ also satisfies (\ref{KR2}), i.e.~it can be approximated by a sum of local terms supported along the boundary of $A$. Intuitively, this is because the above averaging procedure only modifies $Q_A$ near its boundary since the Hamiltonian is commuting and charge conserving. This claim is encapsulated in the following lemma:
\begin{lemma}\label{lemma1}
$K_A^*$ can be approximated, up to $\mathcal{O}(L^{-\infty})$, by a sum of the form
\begin{align}\label{sumr2}
    K_A^{*}&=\sum_{r\in \partial_{\alpha L}A}\overline{K}_{r,A}^{*}+\mathcal{O}(L^{-\infty})   
\end{align}
where $\overline{K}_{r,A}^{*}$ is a strictly local charge conserving operator, with a uniformly bounded norm, supported in $D_{\alpha L}(r)$. 
\end{lemma}


The proof of Lemma~\ref{lemma1} is particularly simple for the special case of strictly local commuting projector Hamiltonians. In fact, in this case, Eq.~(\ref{sumr2}) holds without any error terms. To see this, notice that in (\ref{Kdef}) we only need to include $H_r$ within a finite distance of the boundary of $A$ because all other $H_{r}$'s commute with $Q_A$ exactly. We can then write $K_A^*$ as $K_A^*=\sum_{r\in A}K_{r,A}^*$ where $K_{r,A}^*$ is defined just like $K_A^*$ in (\ref{Kdef}) except with $Q_A$ replaced by $q_r$ and with the averaging restricted to $H_r$'s within a finite distance of the boundary of $A$. Eq.~(\ref{sumr2}) then follows immediately since $K_{r,A}^*$ vanishes exactly except for $r$ within a finite distance of the boundary of $A$. A similar proof holds in the more general case of \emph{almost} local commuting projector Hamiltonians; see the Supplemental Material for details\footnote{See Supplemental Material for the proof of Lemma~\ref{lemma1} for ALCPHs.}.


We now assume Lemma~\ref{lemma1} and proceed with the proof of the theorem. First, we define $K_A^{*-}$ as in (\ref{kmindef}):
\begin{equation}\label{kmindef2}
    K_A^{*-}=\sum_{r\in(\partial_{\alpha L}A)\cap \Lambda^-}\overline{K}_{r,A}^*.
\end{equation}
We then make two observations. The first observation, which follows immediately from the definition (\ref{kmindef2}) and charge conservation, is that
\begin{equation}\label{observ1}
    \mathrm{supp}([K_A^{*-},Q_B])\subset D_{2\alpha L}(r^-),
\end{equation}
where $r^-$ is the point in the lower half torus where the boundaries of $A$ and $B$ intersect (see Fig.~\ref{fig:hallcond}). 
The second observation is that 
\begin{equation}
    [Q_A-K_A^{*-},P_{R^-}] = \mathcal{O}(L^{-\infty}),
    \label{observ2}
\end{equation}
where $R^-=\{r\in\Lambda:r_y< -2\alpha L\}$. To see this, it suffices to show that $[Q_A-K_A^{*-},H_r] = \mathcal{O}(L^{-\infty})$ for any $r \in R^-$ since $P_{R^-}=\prod_{r\in R^-}(1-H_r)$. From (\ref{commH}),  
\begin{align*}
    [Q_A-K_A^{*-},H_r]&=[K_A^{*} - K_A^{*-},H_r].
\end{align*}

The right hand side is $\mathcal{O}(L^{-\infty})$. Indeed, $K_A^{*} - K_A^{*-}$ can be approximated, up to $\mathcal{O}(L^{-\infty})$, by a sum of local terms $\overline{K}_{r,A}^*$ strictly supported in $\{r\in\Lambda:r_y\geq -\alpha L\}$, and each of these $\mathcal{O}(L)$ terms commutes with the almost local terms $H_r$ up to $\mathcal{O}(L^{-\infty})$ according to (\ref{ALCPH3}).

We now use (\ref{observ1},\ref{observ2}) to complete the proof. First, by cyclicity of the trace,
\begin{equation}\label{cyclicity}
\<[Q_A-K_A^{*-},Q_B]\>_{R^-} =\<\{Q_B,[P_{R^-},Q_A-K_A^{*-}]\}\>_{R^-}, 
\end{equation}
where $\{\cdot,\cdot\}$ denotes the anticommutator. By (\ref{observ2}), the right hand side is $\mathcal{O}(L^{-\infty})$; therefore since $[Q_A,Q_B]=0$, we deduce that
\begin{equation}
  \langle[K_A^{*-},Q_B]\rangle_{R^-} =\mathcal{O}(L^{-\infty}). 
  \label{inR2}
\end{equation}
At the same time, using (\ref{wTQO}) together with (\ref{observ1}) and the fact that the distance from $D_{2\alpha L}(r^-)$ to the complement of $R^-$ is proportional to $L$, we have
\begin{equation}
    \langle[K_A^{*-},Q_B]\rangle=\langle[K_A^{*-},Q_B]\rangle_{R^-}+\mathcal{O}(L^{-\infty}).
    \label{wtqo2}
\end{equation}
Theorem~\ref{mainresult} then follows immediately from (\ref{inR2},\ref{wtqo2}). 

It is instructive to compare this proof with the physical argument we presented earlier. To make this comparison, we think of $Q_A - K_A^{*-}$ as the infinitesimal analog of the flux insertion operator $U$. Specifically, we note that the unitary $U$ corresponding to a $2\pi$ flux insertion is given by $U=e^{-2\pi i(Q_A-K_A^{*-})}$\cite{bachmann2020}. We can then see that the two observations (\ref{observ1}, \ref{observ2}) that underlie our proof have close parallels with the physical argument. In particular, (\ref{observ1}) is analogous to our previous claim that $U^\dagger Q_B U - Q_B$ is localized near the point where the support of $U$ intersects $B$. Likewise, (\ref{observ2}) is analogous to our claim that $U$ preserves the ground state away from the punctures. One difference between the two arguments is that the above argument requires that the site Hilbert space is finite dimensional, e.g.~when we cyclically permute the trace in (\ref{cyclicity}), while the physical argument only uses the weaker assumption that $Q_B$ has a bounded spectrum. 

\textbf{\emph{Discussion}}.---While we have focused on Hamiltonians built out of commuting \emph{projectors}, our results apply to a broader class of commuting Hamiltonians. For example, we can replace the projector assumption with a weaker gap assumption: the lowest eigenvalue of $H_r$ is $0$ and it is isolated from the rest of its spectrum by a local gap $g_r\geq g>0$ with $g$ independent of $r,L$. To see why our results apply in this case, note that we can pick a smooth function $\chi_g$ such that $\chi_g(E) = 0$ if $E\leq 0$ and $\chi_g(E) = 1$ if $E\geq g$. We can then spectrally flatten $H_r$ to the projector $\chi_g(H_r)$. Smoothness of $\chi_g$ translates to a rapid decay in real space and so $\chi_g(H_r)$ remains almost local. It follows that $\tilde H = \sum_{r}\chi_g(H_r)$ is an ALCPH with the same ground state space as $H$. Hence, the Hall conductance vanishes for ground states of such almost local, locally gapped, commuting Hamiltonians.

Our results can also be readily extended to \emph{fermionic} systems. Indeed, while the setup for our proof was explicitly bosonic, all results continue to hold in the fermionic setting, provided that we restrict to operators with even fermion parity. This restriction ensures that the locality expressed by (almost) commutation continues to hold in the fermionic setting.

One application of our results is that they provide a short proof that
any system of non-interacting electrons with localized Wannier functions has a vanishing Hall conductance. To prove this, let the projector into the lowest band be $P=\sum_{r,\mu}|\psi_{r,\mu}\rangle\langle\psi_{r,\mu}|$ and let the projector into the other bands be $1-P=\sum_{r,\mu'}|\tilde{\psi}_{r,\mu'}\rangle\langle\tilde{\psi}_{r,\mu'}|$. Here, $\{|\psi_{r,\mu}\rangle\}$ and $\{|\tilde{\psi}_{r,\mu'}\rangle\}$ are pairwise orthogonal, superpolynomially localized Wannier functions. A parent Hamiltonian with lowest band projector $P$ is given by $H=\sum_{r,\mu}a(\psi_{r,\mu})a^\dagger(\psi_{r,\mu})+\sum_{r,\mu'}a^\dagger(\tilde{\psi}_{r,\mu'})a(\tilde{\psi}_{r,\mu'})$, where $a^\dagger(\psi_{r,\mu})$ creates a $\psi_{r,\mu}$ excitation from the Fock vacuum.
%
It is clearly a commuting projector Hamiltonian, and the decay of the Wannier functions implies that the terms are almost localized. Hence $H$ is an ALCPH and our theorem implies that the Hall conductance vanishes. In fact, Ref.~\onlinecite{monaco2018} already proved a stronger version of this result, but our proof has the advantage of applying to a much larger class of interacting systems.


One direction for future work is to investigate which two dimensional topological phases can be realized by ALCPHs in the \emph{absence} of any symmetries. In this case, a reasonable conjecture is that ALCPHs can realize precisely those topological phases that support gapped boundaries. 
Assuming this conjecture, it is particularly interesting to consider topological phases that have a vanishing thermal Hall conductance, but do not support gapped boundaries~\cite{levin2013}. These phases presumably do not have an ALCPH realization, but there is no direct proof of this, to our knowledge.

C.Z. and M.L. acknowledge the support of the Kadanoff Center for Theoretical Physics at the University of Chicago.
This work was supported by the Simons Collaboration on Ultra-Quantum Matter, which is a grant from the 
Simons Foundation (651440, M.L.), and the National Science Foundation Graduate Research Fellowship under Grant No. 1746045. The work of S.B. was supported by NSERC of Canada.
\bibliography{hallcond}

\begin{thebibliography}{23}%
\makeatletter
\providecommand \@ifxundefined [1]{%
 \@ifx{#1\undefined}
}%
\providecommand \@ifnum [1]{%
 \ifnum #1\expandafter \@firstoftwo
 \else \expandafter \@secondoftwo
 \fi
}%
\providecommand \@ifx [1]{%
 \ifx #1\expandafter \@firstoftwo
 \else \expandafter \@secondoftwo
 \fi
}%
\providecommand \natexlab [1]{#1}%
\providecommand \enquote  [1]{``#1''}%
\providecommand \bibnamefont  [1]{#1}%
\providecommand \bibfnamefont [1]{#1}%
\providecommand \citenamefont [1]{#1}%
\providecommand \href@noop [0]{\@secondoftwo}%
\providecommand \href [0]{\begingroup \@sanitize@url \@href}%
\providecommand \@href[1]{\@@startlink{#1}\@@href}%
\providecommand \@@href[1]{\endgroup#1\@@endlink}%
\providecommand \@sanitize@url [0]{\catcode `\\12\catcode `\$12\catcode
  `\&12\catcode `\#12\catcode `\^12\catcode `\_12\catcode `\%12\relax}%
\providecommand \@@startlink[1]{}%
\providecommand \@@endlink[0]{}%
\providecommand \url  [0]{\begingroup\@sanitize@url \@url }%
\providecommand \@url [1]{\endgroup\@href {#1}{\urlprefix }}%
\providecommand \urlprefix  [0]{URL }%
\providecommand \Eprint [0]{\href }%
\providecommand \doibase [0]{http://dx.doi.org/}%
\providecommand \selectlanguage [0]{\@gobble}%
\providecommand \bibinfo  [0]{\@secondoftwo}%
\providecommand \bibfield  [0]{\@secondoftwo}%
\providecommand \translation [1]{[#1]}%
\providecommand \BibitemOpen [0]{}%
\providecommand \bibitemStop [0]{}%
\providecommand \bibitemNoStop [0]{.\EOS\space}%
\providecommand \EOS [0]{\spacefactor3000\relax}%
\providecommand \BibitemShut  [1]{\csname bibitem#1\endcsname}%
\let\auto@bib@innerbib\@empty
\bibitem [{\citenamefont {Kitaev}(2003)}]{Kitaev2003fault}%
  \BibitemOpen
  \bibfield  {author} {\bibinfo {author} {\bibfnamefont {A.~Y.}\ \bibnamefont
  {Kitaev}},\ }\href {\doibase 10.1016/S0003-4916(02)00018-0} {\bibfield
  {journal} {\bibinfo  {journal} {Annals of Physics}\ }\textbf {\bibinfo
  {volume} {303}},\ \bibinfo {pages} {2} (\bibinfo {year} {2003})}\BibitemShut
  {NoStop}%
\bibitem [{\citenamefont {Levin}\ and\ \citenamefont {Wen}(2005)}]{levin2005}%
  \BibitemOpen
  \bibfield  {author} {\bibinfo {author} {\bibfnamefont {M.~A.}\ \bibnamefont
  {Levin}}\ and\ \bibinfo {author} {\bibfnamefont {X.-G.}\ \bibnamefont
  {Wen}},\ }\href {\doibase 10.1103/PhysRevB.71.045110} {\bibfield  {journal}
  {\bibinfo  {journal} {Phys. Rev. B}\ }\textbf {\bibinfo {volume} {71}},\
  \bibinfo {pages} {045110} (\bibinfo {year} {2005})}\BibitemShut {NoStop}%
\bibitem [{\citenamefont {Chen}\ \emph {et~al.}(2013)\citenamefont {Chen},
  \citenamefont {Gu}, \citenamefont {Liu},\ and\ \citenamefont
  {Wen}}]{chen2013}%
  \BibitemOpen
  \bibfield  {author} {\bibinfo {author} {\bibfnamefont {X.}~\bibnamefont
  {Chen}}, \bibinfo {author} {\bibfnamefont {Z.-C.}\ \bibnamefont {Gu}},
  \bibinfo {author} {\bibfnamefont {Z.-X.}\ \bibnamefont {Liu}}, \ and\
  \bibinfo {author} {\bibfnamefont {X.-G.}\ \bibnamefont {Wen}},\ }\href
  {\doibase 10.1103/PhysRevB.87.155114} {\bibfield  {journal} {\bibinfo
  {journal} {Phys. Rev. B}\ }\textbf {\bibinfo {volume} {87}},\ \bibinfo
  {pages} {155114} (\bibinfo {year} {2013})}\BibitemShut {NoStop}%
\bibitem [{\citenamefont {Heinrich}\ \emph {et~al.}(2016)\citenamefont
  {Heinrich}, \citenamefont {Burnell}, \citenamefont {Fidkowski},\ and\
  \citenamefont {Levin}}]{heinrich2016}%
  \BibitemOpen
  \bibfield  {author} {\bibinfo {author} {\bibfnamefont {C.}~\bibnamefont
  {Heinrich}}, \bibinfo {author} {\bibfnamefont {F.}~\bibnamefont {Burnell}},
  \bibinfo {author} {\bibfnamefont {L.}~\bibnamefont {Fidkowski}}, \ and\
  \bibinfo {author} {\bibfnamefont {M.}~\bibnamefont {Levin}},\ }\href
  {\doibase 10.1103/PhysRevB.94.235136} {\bibfield  {journal} {\bibinfo
  {journal} {Phys. Rev. B}\ }\textbf {\bibinfo {volume} {94}},\ \bibinfo
  {pages} {235136} (\bibinfo {year} {2016})}\BibitemShut {NoStop}%
\bibitem [{\citenamefont {Cheng}\ \emph {et~al.}(2017)\citenamefont {Cheng},
  \citenamefont {Gu}, \citenamefont {Jiang},\ and\ \citenamefont
  {Qi}}]{cheng2017}%
  \BibitemOpen
  \bibfield  {author} {\bibinfo {author} {\bibfnamefont {M.}~\bibnamefont
  {Cheng}}, \bibinfo {author} {\bibfnamefont {Z.-C.}\ \bibnamefont {Gu}},
  \bibinfo {author} {\bibfnamefont {S.}~\bibnamefont {Jiang}}, \ and\ \bibinfo
  {author} {\bibfnamefont {Y.}~\bibnamefont {Qi}},\ }\href {\doibase
  10.1103/PhysRevB.96.115107} {\bibfield  {journal} {\bibinfo  {journal} {Phys.
  Rev. B}\ }\textbf {\bibinfo {volume} {96}},\ \bibinfo {pages} {115107}
  (\bibinfo {year} {2017})}\BibitemShut {NoStop}%
\bibitem [{\citenamefont {Haah}(2011)}]{haah2011}%
  \BibitemOpen
  \bibfield  {author} {\bibinfo {author} {\bibfnamefont {J.}~\bibnamefont
  {Haah}},\ }\href {\doibase 10.1103/PhysRevA.83.042330} {\bibfield  {journal}
  {\bibinfo  {journal} {Phys. Rev. A}\ }\textbf {\bibinfo {volume} {83}},\
  \bibinfo {pages} {042330} (\bibinfo {year} {2011})}\BibitemShut {NoStop}%
\bibitem [{Note1()}]{Note1}%
  \BibitemOpen
  \bibinfo {note} {To see why the thermal Hall conductance must vanish for
  commuting Hamiltonians, note that the energy current $f_{jk}=0$ in Eq.~(154)
  of Ref.~\protect \rev@citealp {kitaev2006anyons}, and hence one can choose
  $h_{jkl} = 0$ in Eq.~(159), which leads to a vanishing thermal Hall
  conductance in Eq.~(160).}\BibitemShut {Stop}%
\bibitem [{\citenamefont {Kitaev}(2006)}]{kitaev2006anyons}%
  \BibitemOpen
  \bibfield  {author} {\bibinfo {author} {\bibfnamefont {A.}~\bibnamefont
  {Kitaev}},\ }\href {\doibase 10.1016/j.aop.2005.10.005} {\bibfield  {journal}
  {\bibinfo  {journal} {Annals of Physics}\ }\textbf {\bibinfo {volume}
  {321}},\ \bibinfo {pages} {2} (\bibinfo {year} {2006})}\BibitemShut {NoStop}%
\bibitem [{\citenamefont {Kapustin}\ and\ \citenamefont
  {Fidkowski}(2020)}]{kapustin2020}%
  \BibitemOpen
  \bibfield  {author} {\bibinfo {author} {\bibfnamefont {A.}~\bibnamefont
  {Kapustin}}\ and\ \bibinfo {author} {\bibfnamefont {L.}~\bibnamefont
  {Fidkowski}},\ }\href {\doibase 10.1007/s00220-019-03444-1} {\bibfield
  {journal} {\bibinfo  {journal} {Commun. Math. Phys.}\ }\textbf {\bibinfo
  {volume} {373}},\ \bibinfo {pages} {763} (\bibinfo {year}
  {2020})}\BibitemShut {NoStop}%
\bibitem [{\citenamefont {Bachmann}\ \emph {et~al.}(2020)\citenamefont
  {Bachmann}, \citenamefont {Bols}, \citenamefont {De~Roeck},\ and\
  \citenamefont {Fraas}}]{bachmann2020}%
  \BibitemOpen
  \bibfield  {author} {\bibinfo {author} {\bibfnamefont {S.}~\bibnamefont
  {Bachmann}}, \bibinfo {author} {\bibfnamefont {A.}~\bibnamefont {Bols}},
  \bibinfo {author} {\bibfnamefont {W.}~\bibnamefont {De~Roeck}}, \ and\
  \bibinfo {author} {\bibfnamefont {M.}~\bibnamefont {Fraas}},\ }\href
  {\doibase 10.1007/s00220-019-03537-x} {\bibfield  {journal} {\bibinfo
  {journal} {Commun. Math. Phys.}\ }\textbf {\bibinfo {volume} {375}},\
  \bibinfo {pages} {1249} (\bibinfo {year} {2020})}\BibitemShut {NoStop}%
\bibitem [{\citenamefont {Bachmann}\ \emph {et~al.}(2021)\citenamefont
  {Bachmann}, \citenamefont {Bols}, \citenamefont {De~Roeck},\ and\
  \citenamefont {Fraas}}]{bachmann2021}%
  \BibitemOpen
  \bibfield  {author} {\bibinfo {author} {\bibfnamefont {S.}~\bibnamefont
  {Bachmann}}, \bibinfo {author} {\bibfnamefont {A.}~\bibnamefont {Bols}},
  \bibinfo {author} {\bibfnamefont {W.}~\bibnamefont {De~Roeck}}, \ and\
  \bibinfo {author} {\bibfnamefont {M.}~\bibnamefont {Fraas}},\ }\href
  {\doibase 10.1063/5.0021511} {\bibfield  {journal} {\bibinfo  {journal} {J.
  Math. Phys.}\ }\textbf {\bibinfo {volume} {62}},\ \bibinfo {pages} {011901}
  (\bibinfo {year} {2021})}\BibitemShut {NoStop}%
\bibitem [{\citenamefont {Kapustin}\ and\ \citenamefont
  {Sopenko}(2020)}]{kapustin2020hall}%
  \BibitemOpen
  \bibfield  {author} {\bibinfo {author} {\bibfnamefont {A.}~\bibnamefont
  {Kapustin}}\ and\ \bibinfo {author} {\bibfnamefont {N.}~\bibnamefont
  {Sopenko}},\ }\href@noop {} {\bibfield  {journal} {\bibinfo  {journal} {J.
  Math. Phys.}\ }\textbf {\bibinfo {volume} {61}},\ \bibinfo {pages} {101901}
  (\bibinfo {year} {2020})}\BibitemShut {NoStop}%
\bibitem [{\citenamefont {Laughlin}(1981)}]{laughlin1981quantized}%
  \BibitemOpen
  \bibfield  {author} {\bibinfo {author} {\bibfnamefont {R.~B.}\ \bibnamefont
  {Laughlin}},\ }\href {\doibase 10.1103/PhysRevB.23.5632} {\bibfield
  {journal} {\bibinfo  {journal} {Phys. Rev. B}\ }\textbf {\bibinfo {volume}
  {23}},\ \bibinfo {pages} {5632} (\bibinfo {year} {1981})}\BibitemShut
  {NoStop}%
\bibitem [{\citenamefont {Avron}\ \emph {et~al.}(1994)\citenamefont {Avron},
  \citenamefont {Seiler},\ and\ \citenamefont {Simon}}]{avron1994charge}%
  \BibitemOpen
  \bibfield  {author} {\bibinfo {author} {\bibfnamefont {J.~E.}\ \bibnamefont
  {Avron}}, \bibinfo {author} {\bibfnamefont {R.}~\bibnamefont {Seiler}}, \
  and\ \bibinfo {author} {\bibfnamefont {B.}~\bibnamefont {Simon}},\ }\href
  {\doibase 10.1007/BF02102644} {\bibfield  {journal} {\bibinfo  {journal}
  {Commun. Math. Phys.}\ }\textbf {\bibinfo {volume} {159}},\ \bibinfo {pages}
  {399} (\bibinfo {year} {1994})}\BibitemShut {NoStop}%
\bibitem [{\citenamefont {DeMarco}\ and\ \citenamefont
  {Wen}(2021)}]{demarco2021}%
  \BibitemOpen
  \bibfield  {author} {\bibinfo {author} {\bibfnamefont {M.}~\bibnamefont
  {DeMarco}}\ and\ \bibinfo {author} {\bibfnamefont {X.-G.}\ \bibnamefont
  {Wen}},\ }\href@noop {} {\bibfield  {journal} {\bibinfo  {journal} {arXiv
  preprint arXiv:2102.13057}\ } (\bibinfo {year} {2021})}\BibitemShut {NoStop}%
\bibitem [{\citenamefont {Bravyi}\ \emph {et~al.}(2010)\citenamefont {Bravyi},
  \citenamefont {Hastings},\ and\ \citenamefont {Michalakis}}]{bravyi2010}%
  \BibitemOpen
  \bibfield  {author} {\bibinfo {author} {\bibfnamefont {S.}~\bibnamefont
  {Bravyi}}, \bibinfo {author} {\bibfnamefont {M.~B.}\ \bibnamefont
  {Hastings}}, \ and\ \bibinfo {author} {\bibfnamefont {S.}~\bibnamefont
  {Michalakis}},\ }\href {\doibase 10.1063/1.3490195} {\bibfield  {journal}
  {\bibinfo  {journal} {J. Math. Phys.}\ }\textbf {\bibinfo {volume} {51}},\
  \bibinfo {pages} {093512} (\bibinfo {year} {2010})}\BibitemShut {NoStop}%
\bibitem [{\citenamefont {Michalakis}\ and\ \citenamefont
  {Zwolak}(2013)}]{michalakis2013}%
  \BibitemOpen
  \bibfield  {author} {\bibinfo {author} {\bibfnamefont {S.}~\bibnamefont
  {Michalakis}}\ and\ \bibinfo {author} {\bibfnamefont {J.~P.}\ \bibnamefont
  {Zwolak}},\ }\href
  {https://link.springer.com/article/10.1007/s00220-013-1762-6} {\bibfield
  {journal} {\bibinfo  {journal} {Commun. Math. Phys.}\ }\textbf {\bibinfo
  {volume} {322}},\ \bibinfo {pages} {277} (\bibinfo {year}
  {2013})}\BibitemShut {NoStop}%
\bibitem [{\citenamefont {Hastings}\ and\ \citenamefont
  {Michalakis}(2015)}]{hastings2015}%
  \BibitemOpen
  \bibfield  {author} {\bibinfo {author} {\bibfnamefont {M.~B.}\ \bibnamefont
  {Hastings}}\ and\ \bibinfo {author} {\bibfnamefont {S.}~\bibnamefont
  {Michalakis}},\ }\href {\doibase 10.1007/s00220-014-2167-x} {\bibfield
  {journal} {\bibinfo  {journal} {Commun. Math. Phys.}\ }\textbf {\bibinfo
  {volume} {334}},\ \bibinfo {pages} {433} (\bibinfo {year}
  {2015})}\BibitemShut {NoStop}%
\bibitem [{\citenamefont {Bachmann}\ \emph {et~al.}(2018)\citenamefont
  {Bachmann}, \citenamefont {Bols}, \citenamefont {De~Roeck},\ and\
  \citenamefont {Fraas}}]{bachmann2018}%
  \BibitemOpen
  \bibfield  {author} {\bibinfo {author} {\bibfnamefont {S.}~\bibnamefont
  {Bachmann}}, \bibinfo {author} {\bibfnamefont {A.}~\bibnamefont {Bols}},
  \bibinfo {author} {\bibfnamefont {W.}~\bibnamefont {De~Roeck}}, \ and\
  \bibinfo {author} {\bibfnamefont {M.}~\bibnamefont {Fraas}},\ }\href
  {\doibase 10.1007/s00023-018-0651-0} {\bibfield  {journal} {\bibinfo
  {journal} {Annales H. Poincar{\'e}}\ }\textbf {\bibinfo {volume} {19}},\
  \bibinfo {pages} {695} (\bibinfo {year} {2018})}\BibitemShut {NoStop}%
\bibitem [{\citenamefont {Avron}\ and\ \citenamefont
  {Seiler}(1985)}]{avron1985quantization}%
  \BibitemOpen
  \bibfield  {author} {\bibinfo {author} {\bibfnamefont {J.~E.}\ \bibnamefont
  {Avron}}\ and\ \bibinfo {author} {\bibfnamefont {R.}~\bibnamefont {Seiler}},\
  }\href {\doibase 10.1103/PhysRevLett.54.259} {\bibfield  {journal} {\bibinfo
  {journal} {Phys. Rev. Lett.}\ }\textbf {\bibinfo {volume} {54}},\ \bibinfo
  {pages} {259} (\bibinfo {year} {1985})}\BibitemShut {NoStop}%
\bibitem [{Note2()}]{Note2}%
  \BibitemOpen
  \bibinfo {note} {See Supplemental Material for the proof of Lemma~\ref
  {lemma1} for ALCPHs.}\BibitemShut {Stop}%
\bibitem [{\citenamefont {Monaco}\ \emph {et~al.}(2018)\citenamefont {Monaco},
  \citenamefont {Panati}, \citenamefont {Pisante},\ and\ \citenamefont
  {Teufel}}]{monaco2018}%
  \BibitemOpen
  \bibfield  {author} {\bibinfo {author} {\bibfnamefont {D.}~\bibnamefont
  {Monaco}}, \bibinfo {author} {\bibfnamefont {G.}~\bibnamefont {Panati}},
  \bibinfo {author} {\bibfnamefont {A.}~\bibnamefont {Pisante}}, \ and\
  \bibinfo {author} {\bibfnamefont {S.}~\bibnamefont {Teufel}},\ }\href
  {\doibase 10.1007/s00220-017-3067-7} {\bibfield  {journal} {\bibinfo
  {journal} {Commun. Math. Phys.}\ }\textbf {\bibinfo {volume} {359}},\
  \bibinfo {pages} {61} (\bibinfo {year} {2018})}\BibitemShut {NoStop}%
\bibitem [{\citenamefont {Levin}(2013)}]{levin2013}%
  \BibitemOpen
  \bibfield  {author} {\bibinfo {author} {\bibfnamefont {M.}~\bibnamefont
  {Levin}},\ }\href {\doibase 10.1103/PhysRevX.3.021009} {\bibfield  {journal}
  {\bibinfo  {journal} {Phys. Rev. X}\ }\textbf {\bibinfo {volume} {3}},\
  \bibinfo {pages} {021009} (\bibinfo {year} {2013})}\BibitemShut {NoStop}%
\end{thebibliography}%


%

\end{document}


\title{Supplemental Material}

\author{Carolyn Zhang}
\author{Michael Levin}
\affiliation{Department of Physics, Kadanoff Center for Theoretical Physics, University of Chicago, Chicago, Illinois 60637,  USA}
\author{Sven Bachmann}
\affiliation{Department of Mathematics, The University of British Columbia, Vancouver BC V6T 1Z2, Canada}

\maketitle
In this Supplemental Material, we prove Lemma 1 from the main text. First, we introduce some notation. For any operator $O$ and any subset $S$ of lattice sites, we define a corresponding ``averaged'' operator $\mathbb{E}_S(O)$ by averaging $O$ over the unitaries generated by $\{ H_r : r \in S\}$: 
\begin{align}
    \mathbb{E}_S(O) = \int\!\mathcal{D}_S[\theta] e^{i(\theta,H)_S} O e^{-i(\theta,H)_S}.
\end{align}
where $\mathcal{D}_S[\theta] = \prod_{r\in S}\frac{d\theta_r}{2\pi}$, $(\theta,H)_S = \sum_{r\in S} \theta_r H_r$, and we integrate over $\{0\leq\theta_r\leq 2\pi\}$. In this notation, $K_A^*$ can be written as
\begin{align}
    K_A^* = Q_A - \mathbb{E}_\Lambda(Q_A)
    \label{Kdefsupp}
\end{align}

The averaging operation $\mathbb{E}_S$ has several important properties that we will use below. We note first of all that $\mathbb{E}_S$ is a contraction, i.e. $\Vert \mathbb{E}_S(O)\Vert \leq \Vert O \Vert$, and second of all that, since the Hamiltonian is commuting, $\mathbb{E}_S = \mathbb{E}_{S\setminus S'}\circ \mathbb{E}_{S'} = \mathbb{E}_{S'}\circ \mathbb{E}_{S\setminus S'} $ for any $S'\subset S$ where $S\setminus S'$ is the complement of $S'$ in $S$. It follows in particular that
\begin{align}
\Vert \mathbb{E}_{S}(O) - \mathbb{E}_{S'}(O) \Vert 
\leq \Vert \mathbb{E}_{S\setminus S'}(O) - O \Vert . 
\label{essprime}
\end{align}
%
%
Another useful bound is that
\begin{align}
    \Vert e^{i(\theta,H)_S} O e^{-i(\theta,H)_S} - O \Vert \leq 2\pi \sum_{r \in S} \left\Vert [H_r, O] \right\Vert .
    \label{ineq1}
\end{align}
This bound follows from the fact that 
\begin{equation*}
\left\Vert \frac{d}{dt} (e^{-it(\theta,H)_S} O e^{it(\theta,H)_S}) \right\Vert \leq 2\pi \sum_{r \in S} \left\Vert [H_r, O] \right\Vert,
\end{equation*}
together with the fundamental theorem of calculus. Integrating (\ref{ineq1}) over all $\theta_r$'s gives yet another property of $\mathbb{E}_S$:
\begin{equation}
\Vert \mathbb{E}_{S}(O) - O \Vert \leq 2\pi \sum_{r \in S} \Vert [H_r,O]\Vert.
\label{es}
\end{equation}
Finally, combining (\ref{essprime}, \ref{es}),
we deduce the bound
\begin{equation}\label{AverageFar}
\Vert  \mathbb{E}_{S}(O) - \mathbb{E}_{S'}(O) \Vert \leq 2\pi \sum_{r \in S\setminus S'} \Vert [H_r, O] \Vert .
\end{equation}
%
%

We now apply the above bound (\ref{AverageFar}) to the case $O = Q_A$. In that case, charge conservation guarantees that $[H_r,Q_A]=-[H_r,Q_{A^c}]$ and therefore, by the almost locality of $H_r$,
\begin{equation}\label{edgeR}
[H_r,Q_A] = \mathcal{O}\left(\max(\mathrm{dist}(r,A),\mathrm{dist}(r,A^c))^{-\infty}\right).
\end{equation}
In other words, only $H_r$ with $r$ near the edge of $A$ fail to commute with $Q_A$. Combining this with (\ref{AverageFar}), we see that we can restrict the average in~(\ref{Kdefsupp}) to a strip along the boundary of $A$ with superpolynomially small error:
\begin{equation}
\label{modifiedk}
K_A^{*}=Q_A - \mathbb{E}_{\partial_{\alpha L/2}A}(Q_A)+\mathcal{O}(L^{-\infty}).
\end{equation}

Next, we split $Q_A = \sum_{r\in A}q_r$ in two contributions, $Q_{\mathrm{bdry}}$ and $Q_{\mathrm{int}}$, according to to whether $r\in\partial_{\alpha L}A$ or not. From (\ref{es}), we can see that the contribution to $K_A^{*}$ from $Q_{\mathrm{int}}$ is $\mathcal{O}(L^{-\infty})$ since the distance between $\partial_{\alpha L/2}A$ and $Q_{\mathrm{int}}$ is proportional to $L$. Therefore, we can replace $Q_A$ by $Q_{\mathrm{bdry}}$ in (\ref{modifiedk}):
\begin{equation*}
    K_A^* = Q_{\mathrm{bdry}} - \mathbb{E}_{\partial_{\alpha L/2}A}(Q_{\mathrm{bdry}})+\mathcal{O}(L^{-\infty}).
\end{equation*}

To proceed further, we write $Q_{\mathrm{bdry}}$ as a sum of $q_r$'s and we use (\ref{AverageFar}) to approximate $\mathbb{E}_{\partial_{\alpha L/2}A}(q_r) = \mathbb{E}_{\tilde{D}_{\alpha L/2}(r)}(q_r) + \mathcal{O}(L^{-\infty})$ where $\tilde{D}_{\alpha L/2}(r) = D_{\alpha L/2}(r) \cap \partial_{\alpha L/2}A$. In this way, we derive
\begin{equation*}
     K_A^*= \sum_{r\in(\partial_{\alpha L}A)\cap A}K_{r,A}^*+\mathcal{O}(L^{-\infty})
\end{equation*}
where 
\begin{equation}\label{bounded}
    K_{r,A}^* = q_r - \mathbb{E}_{\tilde{D}_{\alpha L/2}(r)}(q_r)
\end{equation}    
    
All that remains is to show that $K_{r,A}^*$ can be approximated by a strictly local, uniformly bounded, charge conserving operator $\overline{K}_{r,A}^*$ supported within $D_{\alpha L}(r)$. To show this, we need one more property of $\mathbb{E}_S$, namely the inequality
\begin{align}
    \Vert [\mathbb{E}_S(O_1), O_2] \Vert \leq 4 \pi \Vert O_1 \Vert \cdot \sum_{r \in S} \Vert [H_r, O_2] \Vert  
    \label{escomm}
\end{align}
which holds for any two operators $O_1, O_2$ with $[O_1, O_2] = 0$. To derive this inequality, note that
\begin{align}
    \Vert [\mathbb{E}_S(O_1),O_2] \Vert 
   & \leq \int\!\mathcal{D}_S[\theta] \ \Vert [O_1, e^{-i(\theta,H)_S} O_2 e^{i(\theta,H)_S}] \Vert 
    \label{ineq2}
\end{align}
Combining (\ref{ineq1}, \ref{ineq2}) and using $[O_1, O_2] = 0$ gives (\ref{escomm}).

We are now ready to complete the argument and show that $K_{r,A}^*$ is localized near $r$. To do this, we consider the commutator $[K_{r,A}^*, O_x]$ where $O_x$ is an operator supported on some site $x$ outside of $D_{\alpha L}(r)$. Using the above inequality (\ref{escomm}) with $O_1 = q_r$ and $O_2 = O_x$, and $S = \tilde{D}_{\alpha L/2}(r)$, we deduce that
\begin{align}
    \Vert [K_{r,A}^*, O_x] \Vert &\leq 4 \pi \Vert q_r \Vert \cdot \sum_{r \in \tilde{D}_{\alpha L/2}(r)} \Vert [H_r, O_x] \Vert  \nonumber \\
    &\leq \Vert O_x \Vert \cdot \mathcal{O}(L^{-\infty})
    \label{KrAcomm}
\end{align}
where the last line follows from the almost locality of $H_r$.
Since $O_x$ is an arbitrary single site operator supported outside of $D_{\alpha L}(r)$, the bound (\ref{KrAcomm}) implies our claim:  $K_{r,A}^*$ can be approximated up to $\mathcal{O}(L^{-\infty})$ by an observable, $\overline{K}_{r,A}^*$, that is strictly supported inside of $D_{\alpha L}(r)$. [For example, $\overline{K}_{r,A}^*$ can be defined by simply taking the partial trace of $K_{r,A}^*$ over all the lattice sites outside of $D_{\alpha L}(r)$]. It is also clear by construction that these $\overline{K}_{r,A}^*$ operators are charge conserving, and they are uniformly bounded because the norm of $K_{r,A}^*$ is bounded by $2\|q_r\|$ by (\ref{bounded}). This completes the proof of Lemma~1 from the main text.

\bibliography{hallcond}